\documentclass[12pt,preprint]{aastex}
\def\e{{\rm E}}
\def\hbn{{\hfil\break\noindent}}
\begin{document}

\title{Ground-based Microlensing Surveys}
\author{Andrew Gould\altaffilmark{1}, B.\ Scott Gaudi\altaffilmark{1}, and David P.\ Bennett\altaffilmark{2}}
\altaffiltext{1}{Department of Astronomy, The Ohio State University,
140 W.\ 18th Ave., Columbus, OH 43210, USA}
\altaffiltext{2}{Department of Physics, University of Notre Dame, IN 46556, USA}\

\vskip-0.1in
{\bf 1. Overview}

Microlensing is a proven extrasolar planet search method that has
already yielded the detection of four exoplanets.  These detections
have changed our understanding of planet formation ``beyond the
snowline'' by demonstrating that Neptune-mass planets with separations
of several AU are common.  Microlensing is sensitive to planets that
are generally inaccessible to other methods, in particular cool
planets at or beyond the snowline, very low-mass (i.e.\ terrestrial)
planets, planets orbiting low-mass stars, free-floating planets, and
even planets in external galaxies.  Such planets can provide critical
constraints on models of planet formation, and therefore the next
generation of extrasolar planet searches should include an aggressive
and well-funded microlensing component.  When combined with the
results from other complementary surveys, next generation
microlensing surveys can yield an accurate and complete census of the
frequency and properties of planets, and in particular low-mass
terrestrial planets.  Such a census provides a critical input for the
design of direct imaging experiments.  

Microlensing planet searches can be carried out from either the ground
or space. Here we focus on the former, and leave the discussion of
space-based surveys for a separate paper.  We review the microlensing
method and its properties, and then outline the potential of next
generation ground-based microlensing surveys.  Detailed models of such
surveys have already been carried out, and the first steps in
constructing the required network of 1-2m class telescopes with wide
FOV instruments are being taken.  However, these steps are primarily
being taken by other countries, and if the US is to remain
competitive, it must commit resources to microlensing surveys in 
the relatively near future. 

{\bf 2. The Properties of Microlensing Planet Searches}

If a foreground star (``lens'') becomes closely aligned with 
a more distant star (``source''), it bends the source light 
into two images.  The resulting magnification is a monotonic
function of the projected separation. For Galactic stars, the image 
sizes and separations are of order $\mu$as and mas respectively, 
so they are generally
not resolved. Rather ``microlensing events'' are recognized from
their time-variable magnification \citep{pac86}, which typically occurs on
timescales $t_\e$ of months, although it ranges from days to years in
extreme cases.  Presently about 600 microlensing events are
discovered each year, almost all toward the Galactic bulge.

If one of these images passes close to a planetary companion of the lens star,
it further perturbs the image and so changes the magnification.
Because the range of gravitational action scales $\propto \sqrt{M}$, where
$M$ is the mass of the lens, the planetary perturbation typically
lasts $t_p\sim t_\e\sqrt{m_p/M}$, where $m_p$ is the planet mass.
That is, $t_p\sim 1\,$day for Jupiters and $t_p\sim 1.5\,$hours for
Earths.  Hence, planets are discovered by intensive, round-the-clock
photometric monitoring of ongoing microlensing events \citep{mao91,gould92}

{\bf 2.1 Sensitivity of Microlensing}

While, in principle, microlensing can detect planets of any mass and
separation, orbiting stars of any mass and distance from the Sun, 
the characteristics of microlensing favor some
regimes of parameter space.
\hbn $\bullet$ {\bf Sensitivity to Low-mass Planets}:  Compared to other techniques, microlensing
is more sensitive to low-mass planets. This is because the {\it amplitude
of the perturbation does not decline as the planet mass declines},
at least until
mass goes below that of Mars \citep{bennett96}.  The {\it duration} does decline as 
$\sqrt{m_p}$
(so higher cadence is required for small planets) and the probability
of a perturbation also declines as $\sqrt{m_p}$ (so more stars must
be monitored), but if a signal is detected, its magnitude is typically
large (~$\ga 10\%$), and so easily characterized and unambiguous.
\hbn $\bullet$ {\bf Sensitivity to Planets Beyond the Snowline}:  Because microlensing works by
perturbing images, it is most sensitive to planets that lie at projected
distances where the images are the largest.  This so-called
``lensing zone'' lies within a factor of 1.6 of the 
Einstein ring, 
$r_\e = \sqrt{(4 G M/c^2)D_s x(1-x)}$, 
where $x=D_l/D_s$ and  $D_l$ and $D_s$ are the distances to the lens and source. 
At the Einstein ring, the equilibrium temperature is
\begin{equation}
T_\e = T_\oplus\biggl({L\over L_\odot}\biggr)^{1/4}
\biggl({r_\e\over \rm AU}\biggr)^{-1/2}\rightarrow
70\,{\rm K}\,{M\over 0.5\,M_\odot}[4x(1-x)]^{1/4}
\end{equation}
where we have adopted a simple model for lens luminosity $L\propto M^5$,
and assumed $D_s=8\,$kpc.  Hence, microlensing is
primarily sensitive to planets in temperature zones similar to 
Jupiter/Saturn/Uranus/Neptune.  
\hbn $\bullet$ {\bf Sensitivity to Free Floating Planets:}
Because the microlensing effect arises directly from the planet mass, 
the existence of a host star is not required for detection.  
Thus, microlensing maintains significant
sensitivity at arbitrarily large separations, and in particular
is the only method that is sensitive to old, free-floating planets. See
\S~4.
\hbn $\bullet$ {\bf Sensitivity to Planets from 1~kpc to M31}:  Microlensing searches require dense star fields
and so are best carried out against the Galactic bulge, which is 8 kpc away.
Given that the Einstein radius peaks at $x=1/2$, it is most sensitive
to planets that are 4 kpc away, but maintains considerable sensitivity
provided the lens is at least 1 kpc from both the observer and the source.
Hence, microlensing is about equally sensitive to planets in the bulge
and disk of the Milky Way.
However, specialized searches are also sensitive to closer planets
and to planets in other galaxies, particularly M31.  See \S~5.
\hbn $\bullet$ {\bf Sensitivity to Planets Orbiting a Wide Range of Host Stars}:  Microlensing is about equally sensitive to
planets independent of host luminosity, i.e., planets of stars all along 
the main sequence, from G to M, as well as white dwarfs and brown dwarfs.  
By contrast, other techniques are generally challenged to detect planets
around low-luminosity hosts.
\hbn $\bullet$ {\bf Sensitivity to Multiple Planet Systems:} In general, the probability
of detecting two planets (even if they are present) is the square
of the probability of finding one, which means it is usually very small.
However, for high-magnification events, the planet-detection probability
is close to unity \citep{griest98}, and so its square is also near unity \citep{gaudi98}.  
In certain rare cases, microlensing can also detect the moon of a planet \citep{bennett02}.

{\bf 2.2 Planet and Host Star Characterization}

Microlensing fits routinely return the planet/star mass ratio
$q=m_p/M$ and the projected separation in units of the Einstein radius
$b=r_\perp/r_\e$ \citep{gaudi97}.  Historically, it was believed that,
for the majority of microlensing discoveries, it would be difficult to
obtain additional information about the planet or the host star beyond
measurements of $q$ and $b$.  This is because of the well-known
difficulty that the routinely-measured timescale $t_\e$ is a
degenerate combination of $M$, $D_l$, and the velocity of the lens. In
this regime, individual constraints on these parameters must rely on a
Bayesian analysis incorporating priors derived from a Galactic model (e.g.,
\citealt{dong06}).

Experience with the actual detections has demonstrated that the
original view was likely shortsighted, and that one can routinely
expect improved constraints on the mass of the host and planet.  
In three of the four microlensing events yielding exoplanet detections,
the effect of the angular size of the source was imprinted on
the light curve, thus
enabling a measurement of the angular size of the Einstein radius
$\theta_\e=r_\e/D_l$.  This constrains the statistical estimate of $M$
and $D_l$ (and so $m_p$ and $r_\perp$).  In hindsight, one can expect
this to be a generic outcome.  Furthermore, it is now clear that for a
substantial fraction of events, the lens light can be detected during and after
the event, allowing photometric mass and distance estimates, and so
reasonable estimates of $m_p$ and $r_\perp$ \citep{bennett07}.  By
waiting sufficiently long (usually 2 to 20 years) one could use space
telescopes or adaptive optics to see the lens separating from the
source, even if the lens is faint.  Such an analysis has already been
used the constrain the mass of the host star of the first microlensing
planet discovery \citep{bennett06}, and similar constraints for
several of the remaining discoveries are forthcoming.  Finally, in
special cases it may also be possible to obtain information about the
three-dimensional orbits of the discovered planets.

\begin{figure}[ht*]
\epsscale{1.0}
\plottwo{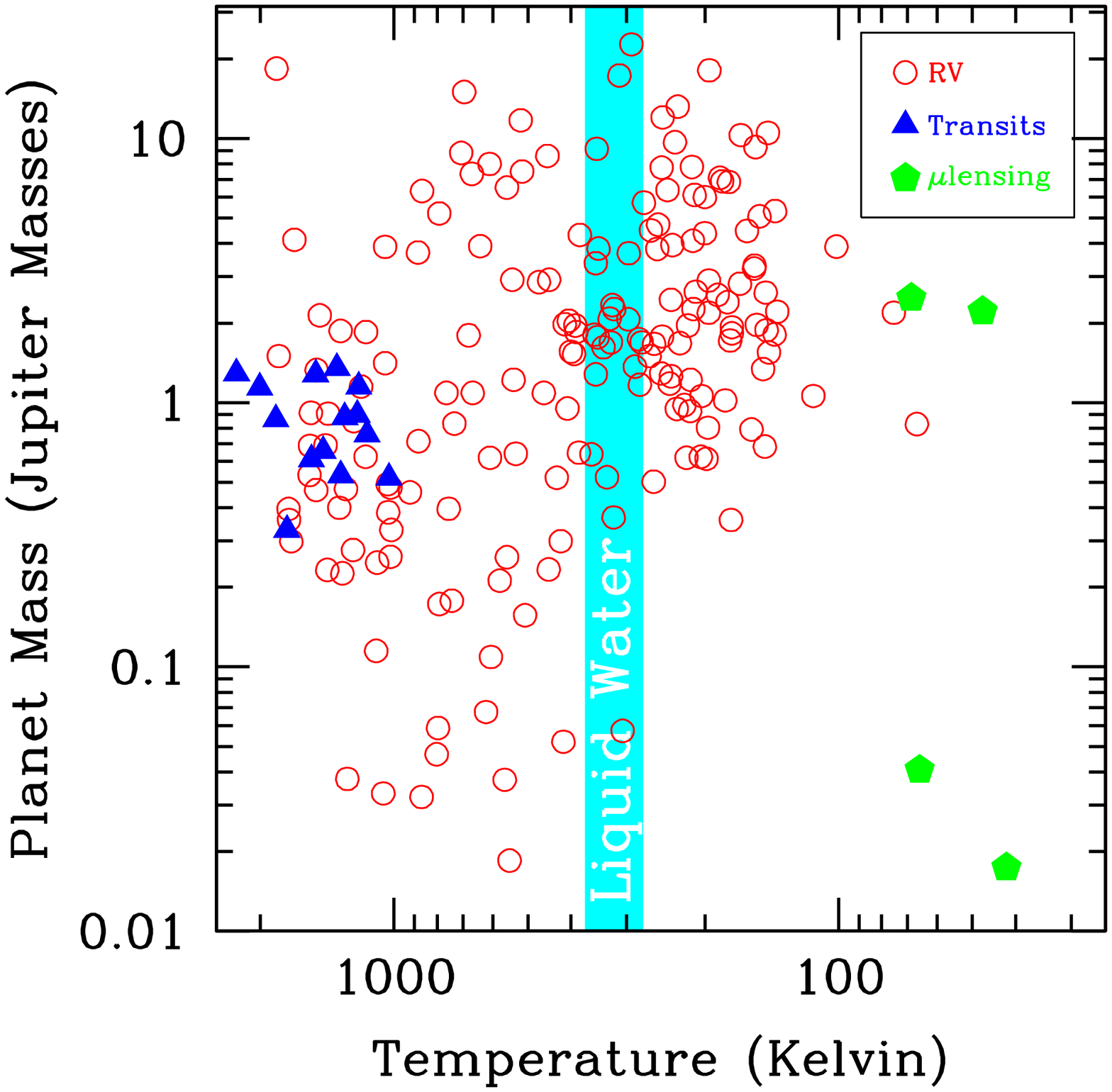}{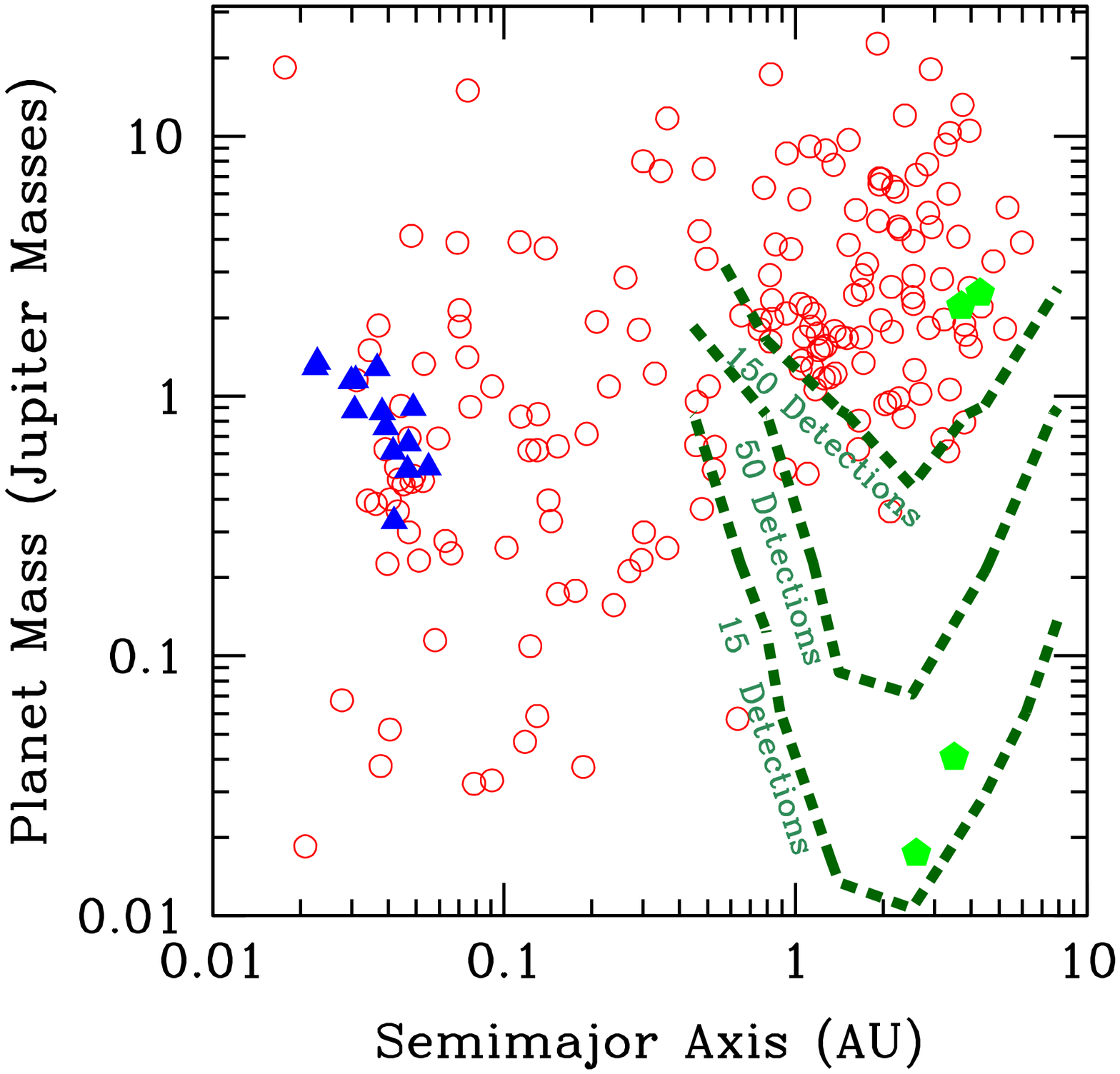}
\caption{
(Left) Known extrasolar planets detected via 
transits (blue), RV (red), and microlensing (green), as a function
of their mass and equilibrium temperature.  (Right) Same as
the right panel, but versus semimajor axis.  The contours show
the number of detections per year from a NextGen microlensing survey. 
}
\end{figure}

\begin{figure}[ht*]
\epsscale{1.0}
\plottwo{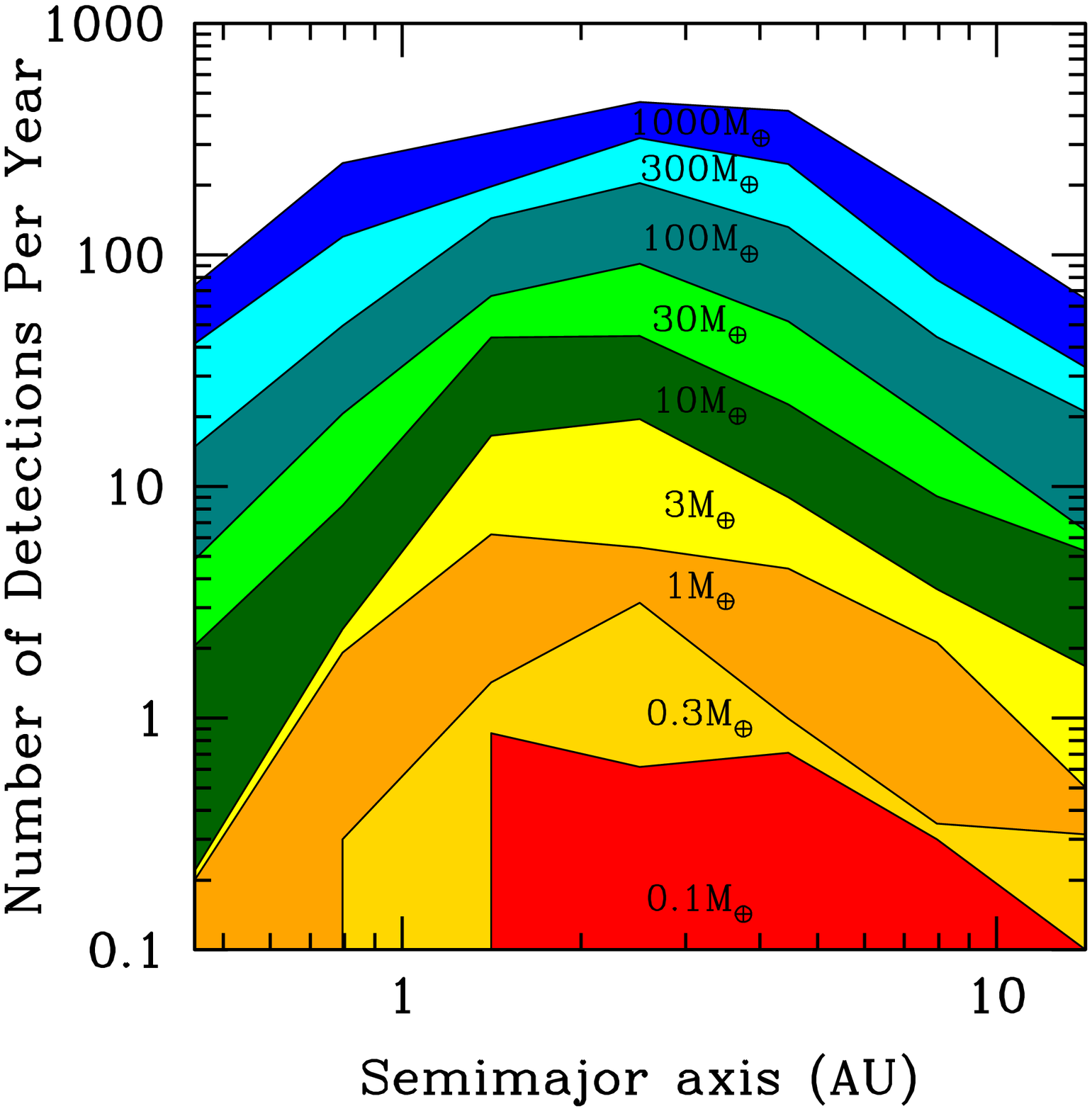}{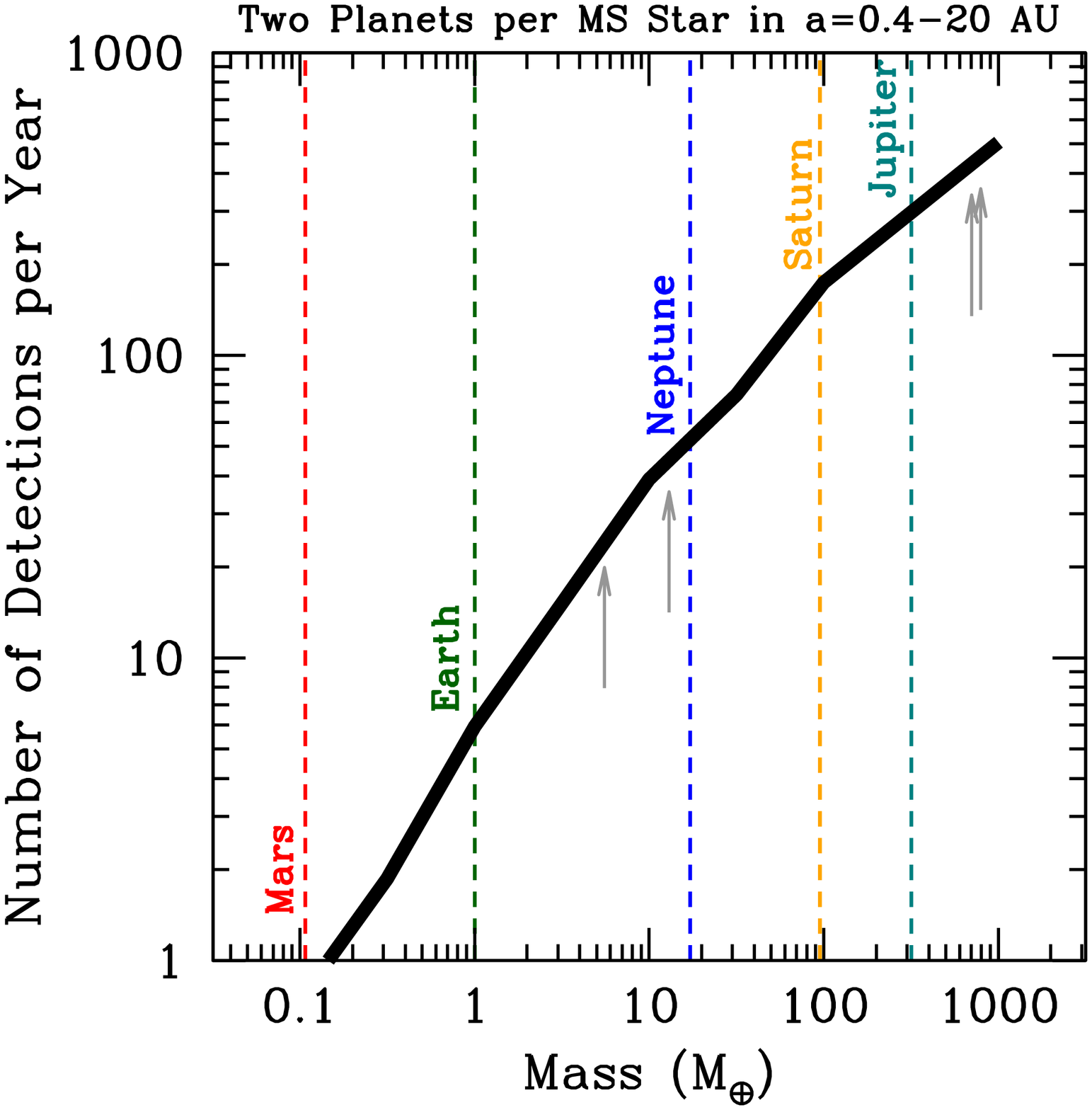}
\caption{
Expectations from a NextGen ground-based microlensing survey.
These results represent the average of two independent simulations
which include very different input assumptions 
but differ in their predictions by only $\sim 0.3$ dex. 
(Left) Number of planets detected per year 
assuming every main-sequence (MS) star has a planet
of a given mass and semi-major axis (see \S 4).  
(Right) Same as left panel, but assuming every MS has two planets distributed
uniformly in $\log(a)$ between 0.4-20~AU. The arrows
 indicate the masses of the four microlensing exoplanet detections.
}\end{figure}

{\bf 3. Present-Day Microlensing Searches}

Microlensing searches today still basically carry out the approach advocated
by \citet{gould92}: Two international networks of astronomers intensively
follow up ongoing microlensing events that are discovered by two other
groups that search for events.  The one major modification is that, following
the suggestion of \citet{griest98}, they try to focus on the
highest magnification events, which are the most sensitive to planets.
Monitoring is done with 1m (and smaller) class telescopes.  Indeed,
because the most sensitive events are highly magnified, amateurs, with
telescopes as small as 0.25m, play a major role.

To date, four secure planets have been detected, all with equilibrium
temperatures $40\,{\rm K}<T<70\,{\rm K}$.
Two are Jupiter
class planets and so are similar to the planets found by RV at these
temperatures \citep{bond04,udalski05}.  However, two are Neptune mass planets, which are
an order of magnitude lighter than planets detected by RV at these
temperatures \citep{beaulieu06,gould06}.  See Figure 1.  This emphasizes the main advantages
that microlensing has over other methods in this parameter range.
The main disadvantage is simply that relatively few planets have
been detected despite a huge amount of work.

{\bf 4. NextGen Microlensing Searches}

Next-generation microlensing experiments will operate on completely
different principles from those at present, which survey large
sections of the Galactic bulge one--few times per night and then
intensively monitor a handful of the events that are identified.  
Instead, wide-field
($\sim 4\,\rm deg^2$) cameras on 2m telescopes on 3--4 continents will 
monitor large ($\sim 10\,\rm deg^2$) areas of the bulge once every 10 minutes
around-the-clock.  The higher cadence will find 6000 events per
year instead of 600.  More important: {\it all 6000 events will 
automatically be monitored for planetary perturbations by the
search survey itself,} as opposed to roughly 50 events monitored per year as
at present.  These two changes will yield a roughly 100-fold
increase in the number of events probed and so in the number
of planetary detections.

Two groups (led respectively by Scott Gaudi and Dave Bennett) have carried 
out detailed simulations of such a survey, taking account of 
variable seeing and weather conditions as well as photometry systematics,
and including a Galactic model that matches all known constraints.
While these two independent simulations differ in detail, they come
to similar conclusions. Figure 1 shows the number of planets
detected assuming all main-sequence stars have a planet of a given mass and given
semi-major axis.  While, of course, all stars do not have planets
at all these different masses, \citet{gould06} have shown that
the two ``cold Neptunes'' detected by microlensing imply that
roughly a third of stars have such planets in the ``lensing zone'',
i.e. the region most sensitive for microlensing searches.
 
Microlensing sensitivity does decline at separations that are larger
than the Einstein radius, but then levels to a plateau, which
remains constant even into the regime of free-floating planets.
In this case, the timescales are similar to those of bound-planet
perturbations (1 day for Jupiters, 1.5 hours for Earths) but there
is no ``primary event''.  Again, typical amplitudes are factor of a few,
which makes them easily recognizable.  If every star ejected $f$ planets
of mass $m_p$, the event rate would be
$\Gamma = 2\times 10^{-5}f\sqrt{m_p/M_j}\,\rm yr^{-1}$ per monitored star.
Since NextGen experiments will monitor 10s of millions of stars for
integrated times of well over a year, this population will easily be
detected unless $f$ is very small. 
Microlensing is the only known way of detecting (old) free-floating planets,
which may be a generic outcome of planet formation \citep{goldreich04,juric07,ford07}.

{\bf 4.1 Transition to Next Generation}

Although NextGen microlensing experiments will work on completely 
different principles, the transition is actually taking place
step by step.  The Japanese/New Zealand group MOA already has
a $2\,\rm deg^2$ camera in place on their 1.8m NZ telescope and
monitors about $4\,\rm deg^2$ every 10 minutes, while covering a much
wider area every hour.  The OGLE team has funds from the Polish government
to replace their current $0.4\,\rm deg^2$ camera on their 1.3m telescope 
in Chile
with a $1.7\,\rm deg^2$ camera.  When finished,
they will also densely monitor several square degrees while monitoring
a much larger area once per night.  Astronomers in Korea and Germany have 
each made comprehensive proposals to their governments to build a major new 
telescope/camera in southern Africa, which would enable virtually
round-the-clock monitoring of several square degrees.  Chinese
astronomers are considering a similar initiative.  In the meantime,
intensive followup of the currently surveyed fields is continuing.

{\bf 5. Other Microlensing Planet Searches}

While microlensing searches are most efficiently carried out toward
the Galactic bulge, there are two other frontiers that microlensing
can broach over the next decade or so.
\hbn $\bullet$ {\bf Extragalactic Planets}:
Microlensing searches of M31 are not presently sensitive to planets,
but could be with relatively minor modifications.
M31's greater distance implies that only more luminous (hence physically
larger) sources can give rise to detectable microlensing events.
To generate substantial magnification, the planetary Einstein ring
must be larger than the source, which generally implies that Jupiters
are detectable, but Neptunes (or Earths) are not \citep{covone00,baltz01}.  Nevertheless,
it is astonishing that extragalactic planets are detectable at all.
To probe for M31 planets, M31 microlensing events must be detected
in real time, and then must trigger intensive followup observations
of the type currently carried out toward the Galactic bulge, but
with larger telescopes \citep{chung06}. This capability is well within reach.
\hbn $\bullet$ {\bf Nearby microlensing events}:
In his seminal paper on microlensing, \citet{einstein36} famously dismissed the
possibility that it would ever be observed because the event rate for the
bright stars visible in his day was too small.  Nevertheless, a Japanese
amateur recently discovered such a ``domestic microlensing event'' (DME)
of a bright $(V\sim 11.4)$, nearby $(\sim 1~{\rm kpc})$ star,  
which was then intensively
monitored by other amateurs (organized by Columbia professor Joe Patterson).
While intensive observations began too late to detect planets,
\citet{gaudi07} showed that more timely observations would have been
sensitive to an Earth-mass planet orbiting the lens.  In contrast to more
distant lenses, DME lenses would usually be subject to followup observations,
including RV.  This would open a new domain in microlensing planet searches.
Virtually all such DMEs could be found with two ``fly's eye'' telescopes, 
one in each hemisphere, which would combine 120 10 cm cameras on a
single mount to simultaneously monitor the $\pi$ steradians above airmass 2
to $V=15$.  A fly's eye telescope would have many other applications
including an all-sky search for transiting planets and a 3-day warning
system for Tunguska-type impactors.  Each would cost $\sim$\$4M.

{\bf 6. Conclusion and Outlook}

In our own solar system, the equilibrium-temperature range probed by
microlensing (out past the ``snow line'') is inhabited by four
planets, two gas giants and two ice giants.  All have similar-sized
ice-rock cores and differ primarily in the amount of gas they have
accreted.  Systematic study of this region around other stars would
test predictive models of planet formation (e.g.\ \citealt{ida04}) by
determining whether smaller cores (incapable of accreting gas) also
form.  Such a survey would give clues as to why cores that reach critical
gas-grabbing size do or do not actually manage to accrete gas, 
and if so, how much.
In the inner parts of this region, RV probes the gas giants but not
the ice giants nor, of course, terrestrial planets.  RV cannot make
reliable measurements in the outer part of this region at all because
the periods are too long. Future astrometry missions (such as {\it
SIM}) could probe the inner regions down to terrestrial masses, but
are also limited by their limited lifetime in the outer regions.
Hence, microlensing is uniquely suited to a comprehensive study of
this region.  

Although microlensing searches have so far detected only
a handful of planets, these have {\it already} changed our
understanding of planet formation ``beyond the snowline''.
Next generation microlensing surveys, which
would be sensitive to dozens of ``cold Earths'' in this region, are
well advanced in design conception and are starting initial practical
implementation.  These surveys
play an additional crucial role as proving grounds for a space-based microlensing survey,
the results of which are likely to completely revolutionize our understanding of
planets over a very broad range of masses, separations, and host star masses
(see the Bennett et al.\ ExoPTF white paper). 

Traditionally, US astronomers have played a major
role in microlensing planet searches.  For example, Bohdan Paczy\'nski
at Princeton essentially founded the entire field \citep{pac86} and
co-started OGLE.  Half a dozen US theorists have all contributed key
ideas and led the analysis of planetary events.  The Ohio State and
Notre Dame groups have played key roles in inaugurating and sustaining
the follow-up teams that made 3 of the 4 microlensing planet
detections possible.

Nevertheless, it must be frankly stated that the field is increasingly
dominated by other countries, often with GDPs that are 5--10\% of the
US GDP, for the simple reason that they are outspending the US by a
substantial margin.  There are simply no programs that would provide
the \$5--\$10M required to be in the NextGen microlensing game.
If US astronomers still are in this game at all, it is because of the
strong intellectual heritage that we bring, augmented by the practical
observing programs that we initiated when the entire subject was being
run on a shoestring.  These historical advantages will quickly
disappear as the next generation of students is trained on NextGen
experiments, somewhere else.

\end{document}